\documentclass[epsfig,12pt]{article}
\usepackage{epsfig,amssymb,euscript,bbold}
\usepackage{amsmath}
\allowdisplaybreaks
\def\be{\begin{equation}}
\def\ee{\end{equation}}
\def\bea{\begin{eqnarray}}
\def\eea{\end{eqnarray}}

\begin{document}

\title{
{\baselineskip -.2in
\vbox{\small\hskip 4in \hbox{hep-th/0601183}}
\vbox{\small\hskip 4in \hbox{DAMTP-2005-122}}
\vbox{\small\hskip 4in \hbox{}}} 
\vskip .4in 
Extremal single-charge small black holes: \\
Entropy function analysis}
\author{Aninda Sinha $^1$
and Nemani V. Suryanarayana $^2$\\
{}\\
{\small{\it $^1$ Department of Applied Mathematics and Theoretical
    Physics,}}\\ 
{\small{\it Wilberforce Road, Cambridge CB3 0WA, U.K.}} \\
{\small{E-mail: {\tt A.Sinha@damtp.cam.ac.uk}}} \\
{\small{$^2$ {\it Perimeter Institute for Theoretical
      Physics,}}}\\ 
{\small{\it 31 Caroline Street North, Waterloo, ON, N2L 2Y5,
    Canada}}\\ 
{\small{E-mail: {\tt vnemani@perimeterinstitute.ca}}}
}
\maketitle
\abstract{We study stretched horizons of the type $AdS_2 \times S^8$
for certain spherically symmetric extremal small black holes in type
IIA carrying only D0-brane charge making use of Sen's entropy function
formalism for higher derivative gravity. A scaling argument is given
to show that the entropy of this class of black holes for large charge
behaves as $\sqrt{|q|}$ where $q$ is the electric charge. The leading
order result arises from IIA string loop corrections. We find that for
solutions to exist the force on a probe D0-brane has to vanish and we
prove that this feature persists to all higher derivative orders. We
comment on the nature of the extremum of these solutions and on the
sub-leading corrections to the entropy. The entropy of other small
black holes related by dualities to our case is also discussed.}

\tableofcontents
\vskip .4cm

\section{Introduction}

Recently there has been progress in our understanding of black holes
which have zero horizon area in supergravity \cite{ad1, rev, sen1,
senpi, entropyfn, senbh} but have non-zero statistical entropy from
independent state counting arguments in string theory. These
geometries are expected to develop non-zero horizon sizes only after
including derivative corrections. Such geometries are referred to as
`small black holes'. For these black holes one uses Wald's entropy
formula \cite{wald} in higher derivative gravity theories.

Further, Sen has developed a formalism to calculate Wald's entropy for
spherically symmetric charged extremal black holes in gravity theories
with abelian gauge fields and real scalar fields \cite{entropyfn} that
have $AdS_2 \times S^D$ as the near horizon geometry. This formalism
allows one to calculate the entropy of such back holes from their near
horizon geometries alone. The simplest extremal small black hole that
has been considered so far is Sen's black hole \cite{sen1} in
heterotic string compactified on $T^5 \times S^1$ which carries two
electric charges. Its microstates come from the degeneracy of states
of a fundamental string with momentum $N$ and winding $W$ around the
$S^1$.

In this note we ask if there are spherically symmetric extremal small
black holes in string theory which carry just one electric charge
under an abelian gauge field. There are hints from AdS/CFT that such
objects should exist in anti-de Sitter spaces (see \cite{nvs} for
example). Here we will be interested in objects in flat space. There
are several ways one can get abelian gauge fields and moduli fields in
string and M-theory compactifications. The simplest are Kaluza-Klein
gauge fields coming from metric and B-fields. Unfortunately, the
relevant bosonic corrections to supergravity actions are not known in
all such cases of interest which makes analyzing this question
difficult. However, if the gauge field and the relevant moduli fields
come from the KK-reduction of the metric components alone, then one
can obtain the relevant bosonic corrections just by dimensional
reduction of the metric dependent terms in the higher dimensional
theory. Luckily in most cases of interest these metric dependent
higher derivative terms are known to leading order. For example, the
1-form RR gauge field and the dilaton of type IIA can be obtained by
dimensional reduction of the 11-dimensional metric of
M-theory. Therefore, as the simplest example we consider type IIA in
10 dimensions and geometries that are electrically charged under the
RR 1-form field and the dilaton turned on.

We use Sen's entropy function formalism \cite{entropyfn} (see also
\cite{kl}) to show that the entropy of such geometries, if non-zero,
will be proportional to the square-root of the electric charge for
large charges. We further note that the value of the dilaton and
therefore the string coupling in the near horizon geometry is also
large for large charges. Thus, one has to consider the strong-coupling
limit, i.e., the M-theory action and the leading $R^4$ corrections
(which are the 1-loop IIA $R^4$ terms). We will use these $R^4$
correction to type IIA supergravity and exhibit a new $AdS_2 \times
S^8$ ($M_3 \times S^8$ where $M_3$ is {\it locally} $AdS_3$) solution
of type IIA (M-theory) to this order which does not exist without the
derivative corrections to the supergravity action. In the spirit of
Sen's entropy function formalism \cite{entropyfn}, postulating that
there exists an extremal black hole solution which interpolates
between this geometry and ${\mathbb R}^{9,1}$, we show that the
entropy of such a small black hole is indeed non-zero to this order.

One can ask what the possible string microstates are that could be
responsible for this entropy of the extremal D0-brane charged small
black hole. Thinking of type IIA as M-theory compactified on $S^1$,
the momenta of point like gravitons are quantised in units of $1/R$
where $R$ is the radius of the compact direction. These gravitons are
1/2-BPS. One can have large degeneracy by considering all possible
multi-particle states all with momenta in the same direction and with
fixed total momentum $q$. The number of these microstates is given by
the number of partitions of $|q|$. A well known generating function is
\begin{equation}
Z(e^{-\beta}) = \prod_{k=1}^\infty \frac{1}{1-e^{-k \, \beta}} =
\sum_{|q|=0}^\infty d_{|q|} ~ e^{- |q| \, \beta}
\end{equation}
where $d_{|q|}$ is the number of partition of the (non-negative)
integer $|q|$. This degeneracy in the saddle point approximation
evaluates to (see \cite{gsw1} for example)
\begin{equation}
d_{|q|} \approx \frac{1}{4 \sqrt{3} ~ |q|} e^{\pi \sqrt{\frac{2
|q|}{3}}}
\end{equation}
for $|q| >>1$. One can associate a statistical entropy $S_{stat} = \ln
\, d_{|q|}$ given by
\begin{equation}
\label{prediction}
S_{stat} \approx \left( \frac{2\pi^2}{3} \right)^{\frac{1}{2}}
\sqrt{|q|}.  
\end{equation}
to these states for large $|q|$. We suggest that these multi-particle
graviton states are responsible for the entropy of the extremal D0
small black hole mentioned above. The M-theory momentum mode from the
type IIA point of view is represented by the extremal D0-brane
solution given by
\begin{eqnarray}
\label{nullsing}
ds^2_{st} &=& -H^{-1/2} dt^2 + H^{1/2} d\vec{x} \cdot d\vec{x} \cr  
e^\phi &=& H^{3/4}, \cr
A &=& (1-H^{-1}) \, dt
\end{eqnarray}
where $\vec{x}$ is a position vector in ${\mathbb R}^9$, $H(\vec{x}) =
1+ |q|/|\vec x|^7$ is a harmonic function over ${\mathbb R}^9$ and $q$
is the electric charge under the RR 1-form gauge field. In this
solution the dilaton diverges as
\begin{equation}
\label{dindzero}
e^{\frac{2\phi}{3}} \sim \sqrt{|q|} ~\,  |\vec{x}|^{-\frac{7}{2}}
~~~ \hbox{as} ~~~ |\vec{x}| \rightarrow 0.
\end{equation}
The geometry has a curvature singularity at $|\vec{x}| = 0$. Further,
as $|\vec{x}| \rightarrow 0$, $g_{tt} \rightarrow 0$ it has a zero
size horizon at $r=0$ making it a null singularity. This is usually
the starting point of various small black holes in the literature. On
general grounds, one expects that such null-singular solutions develop
finite size horizons appropriate for the entropy predicted by some
microstate counting after including derivative corrections. We suggest
that this should happen for the D0-brane solution (\ref{nullsing}) as
well to make it into an extremal black hole with finite size horizon
with entropy in eq.(\ref{prediction}). One expects that the near
horizon limit of the corrected geometry would be an $AdS_2 \times S^8$
according to the definition of extremal black holes in
\cite{entropyfn} with the dilaton stabilising at a large but finite
value. Instead of trying to correct the full geometry of
(\ref{nullsing}) we simply assume here that it does get corrected with
a finite size horizon via the generalised attractor mechanism
\cite{entropyfn}. At this point we should mention the conventional
wisdom of not associating a large (charge dependent) entropy
\cite{mathur} to the D0-brane solution itself assuming that there is a
unique bound state of $|q|$ D0-branes (upto the usual degeneracy of
256 of a short multiplet).

As mentioned earlier we will find that the well known $R^4$
corrections to the 11-dimensional supergravity \cite{atr4,R4M}
dimensionally reduced on $S^1$ to 10 dimensions gives rise to a finite
answer to the entropy. At this order, we find three real solutions and
only one of them gives rise to positive entropy (which accounts for
6.85\% of that in Eq.(\ref{prediction})) and the other two with
negative entropies. One distinguishing feature is that the so-called
Weinhold metric \cite{fgk} in the scalar moduli space for the positive
entropy solution is positive-definite while for the negative solution,
the metric has negative eigen-values. Another curious feature of our
solutions is that the probe D0-brane experiences vanishing force in
these backgrounds. This condition works out to be the same as the
Riemann tensor being covariantly constant for the uplifted
11-dimensional metric. We are unable to ascertain whether our solution
is supersymmetric or not as the relevant supersymmetry transformations
are not known. We comment on this issue in the discussion (for recent
progress in non-supersymmetric attractor mechanism see
\cite{nonsusy}).

The rest of this note is organised as follows. In section 2 we review
the entropy function formalism of Sen. In section 3 we give a general
argument to show that the entropy of the single-charge extremal black
hole of type IIA under consideration is proportional to
$\sqrt{|q|}$ where $q$ is the electric charge. In section 4 we discuss
the objects related to the type IIA one by dualities and reach similar
conclusions. In section 5 we specifically consider the 11-dimensional
$R^4$ correction and calculate the near horizon geometry and the
entropy of the extremal D0-charged small black hole. We also point out
some interesting features of the near horizon solutions we find. We
conclude with a discussion of some consequences, important issues and
open problems in section 6.

\section{The entropy function formalism: A review}

Using Wald's entropy formula for higher derivative gravity in its
extremal limit, Sen \cite{entropyfn} derived an entropy function
$F(u,v, {\bf q}, {\bf p}, {\bf s})$ for a general class of extremal
spherically symmetric black holes having near horizon geometry $AdS_2
\times S^D$ in $D+2$ dimensions. Here $u,v$ stand for the radii of
$AdS_2$ and $S^D$ respectively while ${\bf q,p}$ stand for the
electric and magnetic charges. ${\bf s}$ denotes the near horizon
values for scalar fields in the theory. The extremization of $F$ with
respect to $u$, $v$ and ${\bf s}$ gives the near horizon geometry and
the black hole entropy.

For concreteness, let us review the entropy function formalism at work
for extremal black holes with near-horizon geometry $AdS_2 \times S^2$
in heterotic string theory as introduced by Sen \cite{entropyfn,
senpi}. Here we consider a fundamental string with large momentum and
winding wrapping a circle. For simplicity we compactify the theory on
$T^5\times S^1$ so that we have a black hole in 4-dimensions. The
supergravity action for heterotic string compactified on $T^5
\times S^1$ is given by
\bea
{\cal S}&=&{1\over 32\pi} \int d^4 x \sqrt{-{\rm det} g} \, S{\big (}
R+ G^{\mu\nu}{\partial_\mu S\partial_\nu S\over
  S^2}-G^{\mu\nu}{\partial_\mu T\partial_\nu T\over T^2}\nonumber \\
&+& T^2 G^{\mu\nu} G^{\mu'\nu'}F_{\mu\mu'}^{(1)} F_{\nu\nu'}^{(1)}-
{1\over T^2}G^{\mu\nu} G^{\mu'\nu'}F_{\mu\mu'}^{(2)}
F_{\nu\nu'}^{(2)}{\big )}+{\rm higher ~ derivatives}\,.
\eea
Since extremal black holes are known to have near horizon geometry
$AdS_2 \times S^D$, one looks for a near horizon solution of the form:
\bea
ds^2&=& u(-r^2 dt^2+{dr^2\over r^2})+v d\Omega_{D}^2\\
S&=&s_S\,,\quad T=s_T\,,\quad F_{rt}^{(i)}=e_i\,.
\eea
The starting point for computing Wald's entropy in higher derivative
gravity for spherically symmetric black holes is
\be
S_{BH}=8\pi \int d\Omega {\delta S\over \delta R_{rtrt}}
\sqrt{-g_{rr} g_{tt}}\,,
\ee
where the action is expressed in terms of symmetrized covariant
derivatives of the fields and $R_{\mu\nu\rho\sigma}$ is treated as an
independent variable. The integration is over the angular variables of
$S^D$ the $D$-dimensional spherical horizon. In the near horizon
geometry, the covariant derivative of all fields vanish and it can be
shown that
\be
S_{BH}=F(u,v,{\bf q,p,s})=2\pi (e_i {\partial f\over \partial
  e_i}-f)\,, 
\label{SBH} 
\ee
where $q_i = \partial_{{e_i}} f$ are the electric charges, ${\bf p}$'s
are the magnetic charges and $f$ is given by
\be
f=\int \, d\Omega \sqrt{-{\rm det} g} \,{\cal L}\,,
\ee
where ${\cal L}$ is the lagrangian density. The function $F$ is called
the entropy function and the physical entropy is obtained by
extremizing it with respect to $u$, $v$ and ${\bf s}$. The Legendre
transform in (\ref{SBH}) is only with respect to the electric
charges. There are in principle an infinite number of higher
derivative corrections. Sen \cite{senpi} considers just the
Gauss-Bonnet
\be
L_{R^2}={S\over 16 \pi}
(R_{\mu\nu\rho\sigma}R^{\mu\nu\rho\sigma}-4R_{\mu\nu}
R^{\mu\nu}+R^2)\,.
\ee
This gives
\be
f(u,v,s_S,s_T,e_1,e_2)={1\over 8}s_S uv\left([-{2\over u}+{2\over
    v}+2{s_T^2 e_1^2\over u^2}+2 {e_2^2\over s_T^2 u^2}]-{16\over
  uv}\right)\,,
\ee
where $[~~]$ include the Einstein-Hilbert and $-16/(uv)$ is the
Gauss-Bonnet term. Extremizing $F(u,v,{\bf q}, {\bf s})$ gives the
equations of motion and they admit non-trivial solutions. The positive
entropy solution is given by
\be
(u \rightarrow 8, v \rightarrow 8, s_S \rightarrow \frac{1}{2}
\sqrt{q_1q_2}, s_T \rightarrow {\sqrt{q_1}\over \sqrt{q_2}})\,.
\ee
Here $q_1$ and $q_2$ are proportional to the momentum $N$ and the
winding number $W$ respectively of the fundamental string. The value
of the entropy at this solution is $4\pi \sqrt{N W}$ which exactly
corresponds to that given by the counting of the corresponding BPS
states in string theory.

A feature of the above solution (and the other solutions which were
discarded on physical grounds like signature of $u,v$ for example) is
that the ratio of the Einstein-Hilbert to the Gauss-Bonnet term is
$-1$. This is because $f=EH+R^2=0$ at the solution which follows from
the equation of motion for $S$.

In type II theories a scaling argument shows \cite{senpi} that the
entropy of the corresponding black hole is again proportional to
$\sqrt{NW}$. The counting of string states yields the proportionality
constant to be $2\sqrt{2}\pi$ although there exists no derivation of
this using a higher derivative action.

\section{Entropy of extremal D0 small black hole}

In this section we look at the example of extremal D0 small black hole
of type IIA at a general level. We use Sen's entropy function
formalism reviewed in section 2 to find the general scaling of the
entropy of such a small black hole. Even though we use the
10-dimensional type IIA language the result of this section can be
generalised to any gravity theory in dimension $d\ge 5$ Kaluza-Klein
reduced on a circle.

For this let us start with the following field configurations in type
IIA:
\begin{eqnarray}
\label{twoaconfigs}
ds^2_{AdS_2 \times S^8} &=& s u [ -r^2 \, dt^2 + dr^2/r^2 ]
+ s v [ \sum_{i=0}^4 d\mu_i^2 + \sum_{i=1}^4 \mu_i^2 \,
d\phi_i^2 ], \cr
e^\phi &=& s^{3/2}, ~~~ F_{tr} = e.
\end{eqnarray}
where the metric is taken to be in string frame and $\mu_0 =
cos\theta_1$, $\mu_1 = \sin\theta_1 \cos\theta_2 \cos\theta_3$, $\mu_2
= \sin\theta_2 \cos\theta_2 \sin\theta_3$, $\mu_3 = \sin\theta_1
\sin\theta_2 \cos\theta_4$, $\mu_4 = \sin\theta_1 \sin\theta_2
\sin\theta_4$. We can choose the gauge potential of $F_{tr} = e$ to be
$A_0 = e \, r$. Now recall the 11-dimensional lift formulae:
\begin{eqnarray}
\label{mthlift}
ds_{11}^2 = e^{-\frac{2\phi}{3}} g_{\mu\nu}^{(S)} dx^\mu \, dx^\nu +
e^{\frac{4\phi}{3}} (dx^{(10)} + A_\mu \, dx^\mu)^2.
\end{eqnarray}
Using this the configurations of eq.(\ref{twoaconfigs}) lift to the
pure geometries:
\begin{equation}
\label{elevendconfigs}
ds^2_{11} = u [ -r^2 \, dt^2 + dr^2/r^2 ] +  v \, d\Omega_8^2 + s^2
(dx^{(10)} + e \, r \, dt)^2. 
\end{equation}
Note that the metric in eq.(\ref{elevendconfigs}) amounts to choosing
the frame
\begin{equation}
\label{mixframe}
g_{\mu\nu} = e^{-2\phi/3} g_{\mu\nu}^{(S)} = e^{-\phi/6}
g_{\mu\nu}^{(E)}
\end{equation}
in type IIA in which there is no kinetic term for the dilaton.

\vskip .4cm
\noindent\underline{\bf A general analysis}\\

The first step towards finding the entropy function is to calculate
$f(u,v,e,s)=\int d\Omega_8 \, {\cal L}(u,v,e,s)$. We find it easier to
calculate ${\cal L}(u,v,s,e)$ by substituting the metric in
eq.(\ref{elevendconfigs}) into the 11-dimensional supergravity
lagrangian. This procedure is valid only when the loop corrections of
11-dimensional supergravity are sub-leading. This will happen if the
dilaton $s$ stabilises at very large values. We will see that this is
indeed the case. Further we assume that the 11-dimensional lift ansatz
of eq.(\ref{mthlift}) continues to be valid beyond the supergravity
approximation. This of course is a matter of convention for the field
variables. Let us now find the general features of the function
$f(u,v,s,e)$ and its implications for the entropy function.

For this, note that the 11-dimensional supergravity lagrangian density
is expected to be invariant under reparametrizations of the M-theory
circle: $x_{10} \rightarrow \lambda x_{10}$. This implies 
for the class of metrics in eq.(\ref{elevendconfigs}) that
for the lagrangian density $e$ and $s$ should appear in the
combination $y=es$
and the entropy function $f(u,v,e,s) = \int d\Omega_8 \sqrt{-\det g}
\, {\cal L}$ takes the form
\begin{equation}
\label{guessone}
f(u,v,s,e) = |s| \, g(u,v,se)\,.
\end{equation}
To see this recall that under $x_{10} \rightarrow x'_{10} = \lambda
x_{10}$ the 11-dimensional metric components tranform as
\begin{equation}
g'_{10,10} = \lambda^{-2} g_{10,10}, ~~ g'_{10,m} = \lambda^{-1}
g_{10 m}, ~~ g'_{mn} = g_{mn}
\end{equation}
where $m,n = 0,1, \cdots, 9$. Since $g_{10,10} = e^{\frac{4\phi}{3}}$
and $g_{10m} = e^{\frac{4\phi}{3}} A_m$ and $g_{mn} =
e^{-\frac{2\phi}{3}} g_{mn}^{(S)}$ the action of reparametrizations
reads
\begin{eqnarray}
e^\phi \rightarrow \lambda^{-\frac{3}{2}} e^\phi &\implies&
s\rightarrow \lambda^{-1} s, ~~ \tilde s \rightarrow
\lambda^{-\frac{3}{2}} \tilde s, \cr
F_{tr} \rightarrow \lambda F_{tr} &\implies & e \rightarrow \lambda
e, \cr
e^{-\frac{2\phi}{3}} g_{mn}^{(S)}
\rightarrow e^{-\frac{2\phi}{3}} g_{mn}^{(S)} &\implies& \{u, v\}
\rightarrow \{u,v\}, ~ \{\tilde u, \tilde v \}
\rightarrow \{ \lambda^{-1} \tilde u, \lambda^{-1} \tilde v \}
\end{eqnarray}
where the tilded variables are appropriate for string frame entropy
function (with $\tilde u = u \, s$, $\tilde v = v \, s$ and $\tilde s
= s^{3/2}$). So the invariant combinations are: $\{ u, v, es \}$ in
terms of untilded variables and $\{$ $\tilde u \tilde s^{-2/3}$,
$\tilde v \tilde s^{-2/3}$, $e\tilde s^{2/3}$ $\}$ in terms of tilded
variables. The last piece is the overall $s$ (or $\tilde s$)
dependence multiplying the function $g( u,v, es)$ or $\tilde g(\tilde
u \tilde s^{-2/3}, \tilde v \tilde s^{-2/3}, e\tilde s^{2/3}) $. From
the 11-dimensional point of view this follows from the measure
$\sqrt{-\det g}$ which is proportional to $|s|$ which in turn is
proportional to $\tilde s^{2/3}$. This is again a consequence of
reparametrisations as $d^{11}x \sqrt{-\det g}$ would have to be
invariant under $x_{10} \rightarrow \lambda x_{10}$ too.

Now extremising this function with respect to $s$, $u$ and $v$ give
rise to the following three equations:
\begin{eqnarray}
\label{guesstwo}
\partial_u f = 0 &\implies& \partial_u g = 0, \label{equ} \cr
\partial_v f = 0 &\implies& \partial_v g = 0, \label{eqv}\cr
\partial_s f = 0 &\implies& g + y \partial_y g = 0.\label{eqy} 
\end{eqnarray}
Let us assume that these three equations are independent and admit a
nontrivial solution $u=u_0$, $v=v_0$ and $y=y_0$ which is `universal'
(charge independent). Given these values we can evaluate $g|_{u_0,
v_0, y_0}$ and $\partial_y g|_{u_0, v_0, y_0}$ which are also
universal and generically non-zero (at least $\partial_y g|_{u_0, v_0,
y_0}\ne 0$ requires $g|_{u_0, v_0, y_0} \ne 0$ for $y_0$ to be
non-trivial as assumed). The electric charge $q$ is defined as
$q=\partial_e f$, the canonical conjugate to $e$. This evaluates to:
\begin{eqnarray}
q &=& ~~\, s^2 \, \partial_y g ~~~~ \hbox{for} ~~~~ s>0, \cr
&=& -s^2 \, \partial_y g ~~~~ \hbox{for} ~~~~ s<0.
\end{eqnarray}
For this to make sense we have to have 
\begin{eqnarray}
&& q>0 \iff \partial_y g>0 ~~~ \hbox{or} ~~~~ q<0 \iff \partial_y g
<0 ~~~  \hbox{if} ~~~ s>0, \cr
&& q>0 \iff \partial_y g<0 ~~~ \hbox{or} ~~~~ q<0 \iff \partial_y g
>0 ~~~  \hbox{if} ~~~ s<0.
\end{eqnarray}
Note that we have $|s| = \sqrt{|q|}/\sqrt{|\partial_y g|}$ (compare
this with eq.(\ref{dindzero}) in the D0-brane solution) and therefore
the value of the dilaton in the near horizon geometry is going to be
large for large values of $|q|$ and hence our approach of using the
11-dimensional action is self-consistent. Finally the entropy is
defined as $S_{BH} = 2\pi (e \, q - f)$ which can be rewritten as:
\begin{eqnarray}
\label{guessfour}
S_{BH} &=& 2\pi \, |s| ~ (y \, \partial_y g - g) \cr
&=& 4 \pi \, |s| \, y \,\partial_y g   
= 4 \pi \sqrt{|q|} \, \frac{y \, \partial_y g}{\sqrt{|\partial_y g|}} 
= 4 \pi \sqrt{|q|} \, \left[ \frac{-g}{\sqrt{|\partial_y g|}}\right]
\end{eqnarray}
For $S_{BH}$ to be positive, we require $g<0$ at the solution. The
actual entropy is the value of $S_{BH}$ at its extremum with respect
to $u$, $v$ and $y$. This can be easily seen, using the extremality
conditions in eq.(\ref{guesstwo}), to be:
\begin{equation}
\label{genralentropy}
S_{BH} =  \left(\frac{2\pi^2}{3}\right)^{\frac{1}{2}} \,  K ~
\sqrt{|q|}  
\end{equation}
where the universal coefficient $K$ is
\begin{equation}
\label{univcoeff}
K=- g ~ \sqrt{\frac{24}{|\partial_y g|}}
\end{equation}
This shows that the leading answer to the entropy of the extremal
small black hole carrying an electric charge under the KK-gauge field,
if non-zero, is proportional to the square-root of its charge for
large charge.

This entropy can be re-expressed in terms of the area of the horizon
(volume of the $S^8$ factor) in Einstein frame. For this first recall
the relation between the String and Einstein frame metrics in type IIA
\begin{equation}
g_{\mu\nu}^{(E)} = e^{-\phi/2} g_{\mu\nu}^{(S)} = s^{-3/4}
g_{\mu\nu}^{(S)} 
\end{equation}
So the radius square of $S^8$ in Einstein frame is $v_E = s^{-3/4}v_S
= s^{-3/4} s v = s^{1/4}v$. Therefore area of the horizon in Einstein
frame is
\begin{equation}
\label{harea}
A_H = \hbox{Vol}_{S^8} \, v_E^4 = \hbox{Vol}_{S^8} \, |s| \, v^4
\end{equation}
Using $S_{BH} = 4 \pi |s| \, y \, \partial_y g$ to eliminate $|s|$
from eq.(\ref{harea}) gives the relation
\begin{equation}
S_{BH} = \left[ \frac{4 \pi \, y \, \partial_y g}{\hbox{Vol}_{S^8} \,
v^4} \right] A_H
\end{equation}
The factor in the square brackets is again universal and so we see
that the entropy of our black hole is still proportional to the area
of the horizon. This shows that in this case the fully corrected
leading answer for the entropy still satisfies the `holographic'
principle (i.e, the entropy is proportional to the area of the horizon
divided by the Newton's constant $A_H/G_N$, after restoring the
units). However the sub-leading corrections will, in general, violate
this behaviour.

\section{In other duality frames}

Next we consider the entropies of extremal single-charge small black
holes in other theories related to the one above by dualities. Let us
go through a short chain of dualities starting with M-theory on a
circle with momentum $N$. First compactify the $x^9$-direction and do
a $9-11$ flip. We get type IIA on $S^1$ with total momentum $N$ along
the $x^9$-circle. Performing a T-duality along the $x^9$ circle takes
us to type IIB with a total fundamental string winding charge $N$ on a
dual circle. An S-duality now takes the system to D1-branes wrapped
over $S^1$ in type IIB. Finally, another T-duality along $x^9$ takes
the system back to D0-branes on the circle. It is known that there are
no large extremal black holes carrying the same charges as the ones in
any of the systems in this duality chain related to the D0-brane
system. Assuming that there are going to be extremal small black holes
with the same charges, we would like to ask what their corresponding
entropies are in each of these theories in 9 dimensions.

In the following we argue on general grounds that if such black holes
exist then their entropies (as defined by Sen \cite{wald, entropyfn})
will have to be again given, in each of the duality frames, by $S_{BH}
= \left(2\pi^2/3 \right)^{1/2} ~ K_Q \sqrt{|Q|}$ with unknown
coefficients $K_Q$ as before.

Let us first consider type IIA on $S^1$ with either momentum or
winding charge. In this case, setting $N$ or $W$ to zero in section 2,
apparently implies that the entropy of this black hole is
zero. However, we notice that the scaling argument leading to the
$\sqrt{NW}$ dependence in the entropy formula only involved string
tree-level corrections. The 10-dimensional dilaton stabilises at
$1/\sqrt{W}$ which blows up when $W=0$ and hence in this case the
analysis breaks down and the string loop corrections become
important. The size of the compact circle in the string frame
stabilises at $\sqrt{N/W}$ and for a sensible higher derivative
expansion we want this to be large. Thus, if $N=0$ then the higher
derivative expansion breaks down. In both these cases the scaling
argument needs to be modified.

%
%
%
%

To modify the calculation correctly, start with type IIA and choose the
Einstein frame. As before let us assume that we can obtain the
9-dimensional action from the 10-dimensional one by the Kaluza-Klein
reduction ansatz
\begin{equation}
ds^2_9 = g_{\mu\nu}(x) \, dx^\mu \, dx^\nu + T^2(x) \, (d\xi +
A^{(1)}_\mu dx^\mu)^2. 
\end{equation}
For a momentum black hole we again have only the metric to worry about
in 10 dimensions. Then arguments similar to the ones used in M-theory
calculation of section 3 show that the relevant $f$ function takes the
following general form:
\begin{equation}
f(u, v, u_S=s, u_T=t, e_1) = t \, g(u, v, s, y = t e_1)
\end{equation}
where $|T(x)| = t$. It is clear that a similar calculation to the
one in section 3 gives rise to an entropy
\begin{equation}
S_{BH} (N) = \left( \frac{2\pi^2}{3} \right)^{\frac{1}{2}} K_N
\sqrt{|N|} 
\end{equation}
where $K_N$ is again an unknown (but universal number). Note that for
the momentum case the radius $t$ of the compact direction in the near
horizon geometry stabilises at only very large values. So
\begin{equation}
\label{moexp}
f (u, v, s, t, e_1) = \sum_{n=0}^\infty t^{1-n} \, g^{(n)}(u, v, s,
e_1 t) + \cdots 
\end{equation}
is a valid expansion with the expansion parameter being 1/t and the
leading term being the usual $t \, g^{(0)} (u, v, s, e_1 t)$. The
``$\cdots$'' represent terms of the type $e^{-t}\, g(u,v,s,e_1t)$ etc
if they exist (see section 6 for more comments on these terms).

Next suppose we have winding strings with charge $W$ along the same
circle. Then in this case one expects that the circle radius $\tilde
t$ stabilises at $\tilde t << 1$. Then the natural expansion would
have to be around $\tilde t << 1$ in powers of $\tilde t$. That is
\begin{equation}
\label{wiexp}
\tilde f (\tilde u, \tilde v, \tilde s, \tilde t, e_2) =
\sum_{n=0}^\infty \tilde t^{n-1} \,\tilde g^{(n)} (\tilde 
u, \tilde v, \tilde s, \frac{e_2}{\tilde t}) + \cdots 
\end{equation}
with the first term contributing to the leading answer of the
entropy. It is easy to see that if we have $\tilde f (\tilde u, \tilde
v, \tilde s, e_2,\tilde t) = (1/\tilde t)\, \tilde g^{(0)} (\tilde u,
\tilde v, \tilde s, e_2/\tilde t)$ then we again get $S_{BH} (W) = (2
\pi^2/3)^{1/2} \tilde K_W \sqrt{|W|}$. Unfortunately this is not the
natural expansion one expects in supergravity with $e_2$ coming from
the KK reduction of $B_{9\mu}$. So a priori it is not at all clear
whether the entropy function $f$ takes the form $\tilde f = (1/\tilde
t) \, \tilde g^{(0)} (\tilde u, \tilde v, \tilde s, e_2/\tilde t) +
\cdots$.

However such an expansion is indeed expected for the following
reason. When $\tilde t << 1$ then the winding modes become very
important and the usual supergravity breaks down. For this case it is
natural to work with the T-dual variables where $\tilde t$ is replaced
by $1/t$ and $e_1$ by $e_2$. That is, the KK gauge field obtained by
dimensional reduction of the B-field is interchanged with that of the
T-dual metric on the dual circle. However we already know that for
this case we have an expansion of the type eq.(\ref{moexp}). Hence we
can start with eq.(\ref{moexp}) and do a T-duality transformation term
by term to get to eq.(\ref{wiexp}). The existence of (\ref{moexp}) in
type IIA on $S^1$ (type IIB on $S^1$) implies the existence of
(\ref{wiexp}) in type IIB on the dual $S^1$ (type IIA on the dual
$S^1$).

Hence we conclude that, if we have $\sqrt{|N|}$ type entropy for both
type IIA and type IIB on a circle with momentum $N$, then it follows
that the entropy of winding charge extremal small black hole (if
non-zero) is proportional to $\sqrt{|W|}$. The entropy of the winding
charge black hole might be understood again by considering
multi-string states with total winding number distributed among
several strings. This is again a problem of partitions of $W$ and so
will give the statistical entropy $S_{BH} = (2\pi^2/3)^{1/2}
\sqrt{|W|}$ for $|W| >> 1$. 

Similar arguments exist in the remaining two duality frames (D1-branes
on $S^1$ in type IIB and D0-branes on $S^1$ in type IIA) to show that
if the corresponding extremal small black holes have non-vanishing
entropies then they would have to be proportional to square-root of
their corresponding electric charges in 9 dimensions.


\section{Entropy of extremal D0 small black hole: Specifics}

In this section we return to the D0-charged small black hole that was
considered in section 2. Here we will calculate the entropy function
by including the 1-loop $R^4$ corrections of type IIA (equivalently
the leading $R^4$ corrections to M-theory).

\subsection{The Einstein-Hilbert term}

The Ricci scalar of the metrics in eq.(\ref{elevendconfigs}) evaluates
to:
\begin{equation}
R=\frac{e^2 s^2}{2u^2} -\frac{2}{u} + \frac{56}{v}. 
\end{equation}
Recall $\hbox{Vol} (S^{n-1}) = 2\pi^{n/2}/\Gamma(n/2)$ implies
$\hbox{Vol}_{S^8} = 32\pi^4/105$. The Einstein-Hilbert term integrated
on $S^8$ then becomes
\begin{equation}
\label{zordrentrpyfn}
f(u,v,s,e) = \hbox{Vol}_{S^8} ~ u \, v^4 \, s \left[\frac{e^2
s^2}{2u^2} -\frac{2}{u} + \frac{56}{v}\right]. 
\end{equation}
Before going on we need to check that this matches precisely with the
type IIA action evaluated on configurations in (\ref{twoaconfigs}). In
this case we get $\tilde s^{-2} \tilde u \tilde v^4 (56/\tilde v -
2/\tilde u)$ from the EH term (in string frame) and $\tilde u \tilde
v^4 e^2/(2\tilde u^2)$ from the gauge field kinetic term where $\tilde
u = s u$, $\tilde v = s v$ and $\tilde s = s^{3/2}$. There is no
contribution from the kinetic term of the dilaton. Using these it is
easy to see that the entropy function (\ref{zordrentrpyfn}) is
reproduced.

\subsection{$R^4$ corrections in $M$-theory}

$R^4$ corrections in $M$-theory have been discussed in several
places \cite{R4M,atr4}. We will follow the notation in \cite{atr4}. In
$M$-theory, there are two sources of $R^4$ terms: $(i)$ $(t_8
t_8-{1\over 4} \epsilon_8 \epsilon_8) R^4$ arising from the one-loop
$R^4$ terms in IIA. $(ii)$ $\epsilon_{11} C_3 R^4$ arising from the
well-known $C_4\wedge X_8$ term in $M$-theory. It is convenient to
split the contributions into 2 invariants:
\be
J_0=t_8 t_8 R^4+{1\over 4} E_8 \,,\quad {\cal I}_2={1\over 4} E_8+2
\epsilon_{11} C_3 [{\rm tr}R^4-{1\over 4}( {\rm tr} R^2)^2]\,,
\ee
where $E_8={1\over 3!} \epsilon_{11}\epsilon_{11} R^4$. Since in our
case we will have $C_3=0$, in terms of
these invariants, the 11-dimensional action can be written as
\be
S=S_0+S_1\,,
\ee
where
\be
S_0={1\over 2 \kappa_{11}^2} \int d^{11}x \sqrt{g} R\,,
\ee
and
\be
\label{sone}
S_1=c \int d^{11}x \sqrt{g} (J_0-2{\cal I}_2)\,.
\ee
Here $c={1\over (2\pi)^4 3^2 2^{13}} (2\pi)^{2/3}
(2\kappa_{11}^2)^{-1/3}$.

In any higher derivative gravity theory there is a field-redefinition
ambiguity (see for example \cite{atfr}). This arises from the fact
that any non-singular field-redefinition will leave the S-matrix
invariant. As a result, the coefficients of terms in the
higher derivative action which can be adjusted through such
field-redefinition are potentially ambiguous. Imposing additional
``off-shell'' constraints can remove this ambiguity. Such
``off-shell'' constraints include imposing absence of propagator
correction and ``off-shell'' supersymmetry. However, such constraints
are hard to find and in general one chooses a suitable scheme to
simplify the higher derivative terms. An example of such a scheme is
to remove the terms proportional to the Ricci tensor by using the
lowest order equations of motion. Specifically for the vacuum
solution,
\be
R_{\mu\nu}-{1\over 2} g_{\mu\nu}R=O(\kappa_{11}^{4/3})\,.
\ee
Thus in the higher derivative $R^4$, terms proportional to the Ricci
tensor will be $O(\kappa_{11}^{8/3})$ and hence will contribute only
at a higher order. Using this scheme, it can be shown \cite{atr4} that 
\be
J_0=3\cdot 2^8
\left(R^{hmnk}R_{pmnq}R_h^{\phantom{x}rsp}R^q_{\phantom{q}rsk}+{1\over
  2}R^{hkmn}R_{pqmn}R_h^{\phantom{x}rsp}R^q_{\phantom{q}rsk}\right)\,.
\ee
Furthermore \cite{atr4},
\be
E_8(M_3 \times S^8)=E_8 (S^8)+ a E_2(M_3) E_6(S^8)\,,
\ee
where 
\be
E_{2n}(M^d)=\epsilon_d \epsilon_d R^n\,, d\geq 2n\,.
\ee
The contributing $E_8$ term is explicitly
\be
E_8(S^8)=-\epsilon_{s_1s_2\cdots s_7s_8}\epsilon^{t_1t_2\cdots t_7t_8}
R^{s_1s_2}_{\phantom{s_1s_2}t_1 t_2}\cdots
R^{s_7s_8}_{\phantom{s_7s_8}t_7t_8}\,,
\ee
where $s_i,t_i$ are coordinates on $S^8$ and the minus sign originates
from the Lorentzian signature of the metric on $M_3$.

\subsection{The entropy function: Results}

Evaluating $J_0$, $E_8$ for the metrics in eq.(\ref{elevendconfigs})
yields the following \bea J_0&=& (6{y^4 (5y^2-8u)^2\over
u^8}+{b_1\over v^4})\,,\\ E_8&=&- 8! \, \cdot 2^4 \left[\frac{1}{v^4}
+ \frac{y^2-4u}{4u^2v^3} \right] \,, \eea where $y=se$ and $b_1=3\cdot
2^8 \cdot 1008$.

This yields the reduced entropy function $g(u,v,y=se)$ to be
\be
g(u,v,y)={\rm Vol_{S^8}} u v^4 \left(2({y^2\over 4 u^2}-{1\over u}
+{28\over v})+ c(J_0-{1\over 2} E_8) \right)\,.
\ee
We have chosen $2 \kappa_{11}^2=1$. Note that if
$E_8$ was absent then 
following the general arguments{\footnote{Let us examine some general
forms of the entropy function for 
solutions with positive entropy to exist. The Ricci-scalar works out
to be $a y^2/(4 u^2)-1/u+b/v$ for $M_3 \times S^D$. Let us assume
that the function $g(u,v,y)$ takes the following general form
\be
g(u,v,y)= C \,u\, v^n \left(a {y^2\over 4 u^2}-{1\over u}+{b\over
  v}+f_1(u,y)+f_2(u,v,y)+f_3(v)\right)=C\, u\, v^n h(u,v,y)\,.
\ee
Here $C,a,b$ are positive constants.
We see that for $g<0$ to be satisfied, $h<0$ at the
solution, assuming $u,v>0$. Now we see that for (\ref{eqv}) to be
satisfied, 
\be
n\, u \,v^{n-1} h + u\, v^n\, ({-b \over v^2}+\partial_v (f_2+f_3))=0\,.
\ee
From here we see that if $f_2=0$ and $\partial_v f_3<0$ then $h>0$ and
hence a solution cannot exist. We will find that this will be the case
for $M$-theory on $T^n$ for $n\geq 3$ at $O(R^4)$. Furthermore, it can
also be easily shown that if the higher derivative terms led to a
function of $u$ or a function of $y$ only, then there would be no
solution with positive entropy either.}} for the existence of
solutions, we would conclude the absence of any positive entropy
solution to this order. This will happen for instance when we
compactify on $T^n$ with $n\geq 3$. Evaluating solutions to the
reduced attractor type equations (\ref{guesstwo}) can be achieved
numerically on Mathematica. This yields only one solution with real
positive entropy
\bea u_0&=& 0.0061\\ v_0&=& 0.0559\\ y_0&=& \pm 0.0779\\ \eea with \be
K=0.0685 \,.  \ee The two signs of $y_0$ correspond to solutions with
positive and negative charge $q$. This shows that the higher
derivative corrections have stretched the horizon to a finite size
with $6.85\%$ of the expected entropy answer in
eq.(\ref{prediction}). Of course in order to produce the exact result,
one might need to incorporate all the higher derivative
corrections. Using the above solution $s^2\sim q$, so that for $q>>1$,
we are in the strong coupling regime. This justifies the neglect of
winding mode corrections which produce the tree level IIA $R^4$ term.

In addition to the above solution there are other real solutions which
provide negative entropy. These are:
\be
(u_0,v_0,y_0)=(0.1472,0.6144,\pm 0.1213),\, (0.0085,-0.0816,\pm
0.0923)\,,
\ee
{with $K$ values $-1.97$ and $-0.2067$ respectively.}

A characteristic feature, similar to the heterotic case where the
ratio of the Einstein-Hilbert to the $R^2$ term was $-1$, is that in
this case $EH/R^4=-1/3$. We will explain this observation in the next
section.

\subsection{$u=y^2$ for real solutions}

In the above mentioned results which were obtained numerically on
Mathematica, we find a peculiar feature. In all the real solutions,
$u=y^2$. In fact $(g+y\partial_y g)|_{u=y^2}=u \partial_u g|_{u=y^2}$
and hence even analytic solutions probably exist for $(u,v,y)$.
%
%
%
%
The condition $u=y^2$ also implies that the force on a probe
$D0$-brane vanishes to leading order. In order to see this we write
the $D0$-brane action as
\be
\int dt \, e^{-\phi} \sqrt{g_{00}\partial_0 X^0 \partial_0
  X^0+g_{mn}\partial_0 X^m \partial_0 X^n}-\int A_0 \, dt\,,
\ee
where we have chosen static gauge. In terms of the M-theory
coordinates, $g_{00}=e^{2\phi/3} u r^2$ so that the potential from the
DBI part is $e^{-2\phi/3}r \sqrt{u}= \sqrt{u}r/s$ whereas $\int A_0 \,
dt=\int dt ~ e\, r$.

Thus the condition for no-force is $\sqrt{u}=e s=y$. This condition
also implies that the covariant derivative of the Riemann tensor of
the 11-dimensional metric vanishes. To see this notice that the
non-zero independent components of the Riemann tensor with indices on
$M_3$ are
\begin{equation}
R_{trtr} = \frac{1}{4}(3y^2 + \frac{y^2}{u} - 4u), ~ R_{trtx} = -
\frac{sy^3}{4ru}, ~ R_{txtx} = - \frac{s^2 y^2 r^2}{4u}, ~ R_{rxrx} = 
\frac{y^2 s^2}{4ur^2}
\end{equation}
The non-vanishing independent components of $\nabla_m R_{ijkl}$ are:
\begin{equation}
\nabla_t R_{trtx} = \frac{syr^2}{2u} (y^2-u), ~ \nabla_r R_{trtr} =
\frac{y^2}{u r} (y^2-u), ~ \nabla_r R_{trtx} = \frac{sy}{2ur^2} (u-y^2)
\end{equation}
So the covariant derivative of the Riemann tensor of $M_3$ vanishes
only when $u=y^2$.

Finally let us note some properties of our solutions in the
11-dimensional theory. Firstly, they have 1-dimensional
null-boundaries. To see this rewrite the metric using $z = 1/r$ in
(\ref{elevendconfigs}) as
\begin{equation}
ds^2_{11} = \frac{1}{z^2} \left[ u (-dt^2+dz^2) + vz^2\, d\Omega_8^2 + 
  s^2 z^2 (dx^{(10)} + \frac{e}{z} dt)^2 \right]
\end{equation}
Taking the limit of $z\rightarrow 0$ of $z^2 \, ds_{11}^2$ we get
$-(u-e^2s^2) dt^2$ as the metric on the conformal boundary of $M_3
\times S^8$. Since $u=y^2$ on our solutions we conclude that the
11-dimensional geometries have null 1-dimensional boundaries. Further,
one can write $R_{ab}^{~~~cd} = - (4u)^{-1} (\delta_a^{~c}
\delta_b^{~d} - \delta_a^{~d} \delta_b^{~c})$ for the solutions. So we
conclude that $M_3$ is {\it locally} $AdS_3$ and therefore has a
constant negative scalar curvature $-3/(2 \,u)$ and covariantly
constant Riemann tensor. These solutions are not supported by any
gauge field fluxes and they would not have existed in 11-dimensional
supergravity without the derivative corrections.

As stated in the previous section, for our solutions,
$EH/R^4=-1/3$. After some manipulations it can be shown that
\be
(u\, \partial_u g+{2\over 3} \,v\, \partial_v g)|_{u=y^2}=0=3 EH+R^4\,,
\ee
where the first equality uses the equations of motion. The result
$EH/R^4=-1/3$ immediately follows.

\subsection{Proof of zero-force condition to all orders in
  $\alpha'$}

In this section we will provide a proof of the fact that $u=y^2$ will
lead to two of the equations of motion becoming equal to all orders in
$\alpha'$. We will call this the zero-force condition. Let us first write
\be
g= u f\,,
\ee
where
\be
f=\sum_{n,m=0}^{\infty} c_{nm} ({y\over u})^{2n} ({1\over u})^m\,,
\ee
where $c_{nm}$ could be constants or functions of $v$.
Then
\bea
u\partial_u g=0 &\Rightarrow &  u f-u\sum_{n,m=0}^{\infty} c_{nm}
{y^{2n}\over u^{2n+m}}(2n+m)=0 \\
g+y \partial_y g=0 &\Rightarrow & u f+u \sum_{n,m=0}^{\infty} c_{nm}
{y^{2n}\over u^{2n+m}} (2n)=0\,.
\eea
Thus the two equations above becoming equal when $u=y^2$ leads to 
\be
\sum_{n,m=0}^{\infty} (4n+m) {c_{nm}\over u^{n+m}}=0\,.
\ee
With $m+n=k$ fixed, this leads to
\be
\sum_{n=0}^k (3n+k) c_{n,k-n} =0\,. \label{condint}
\ee
Let us demonstrate that $R_{ab}R^{ab}$ satisfies the above
criterion. Here and in what follows we will focus only on the
3-dimensional part of the space-time $M_3 \times S^8$. Now
$R_{ab}R^{ab}=(3y^4-8 y^2 u+8 u^2)/(4u^4)$. Thus we 
identify $c_{02}=2, c_{11}=-2, c_{20}=3/4$. On the other hand the
condition (\ref{condint}) leads to $2 c_{02}+5 c_{11}+8 c_{20}=0$
which is obviously satisfied with the above $c_{nm}$. In a
straightforward manner $R^2$ also can be showed to satisfy the
zero-force condition. Further suppose we can write $g=u f_1 f_2$ such
that $f_1$ and $f_2$ individually satisfied the zero-force
condition. Then it is easy to see using the chain-rule that $g$ will
also satisfy the zero-force condition. This leads us to conclude that
if the entropy function can be shown to be a function of the
3-dimensional Ricci
scalar and the 3-dimensional Ricci tensor, then the function will
satisfy the 
zero-force condition.
This is straightforward once we notice that
\be
R_{abcd}=-{1\over 2}
g_{ac}g_{bd}R+g_{ac}R_{bd}+g_{bd}R_{ac}-c\leftrightarrow d\,.
\ee
This means that
higher derivative corrections involving the Riemann tensor will be
functions of the Ricci scalar and Ricci tensor only. This completes
the proof.

\subsection{Nature of extremum}

It is an interesting and important question to study the conditions
when the entropy function will be a maximum or a minimum. 
%
%
After some algebra it can be shown that
\be
S_{BH}''=\pi \sqrt{q} {g''^2\over g'^{3/2}}(y+2 {g'\over g''})\,,
\ee
where $'$ denotes differentiation with respect to $y$.
Here we have used the e.o.m to set the term proportional to $(yg'+g)$
in the derivative to zero. Thus we have $y+2 {g'\over g''}<0$ ($y+2
{g'\over g''}>0$) for maximum (minimum). Using the equations of motion
we can rewrite this as $gg''>2 g'^2$ ($gg''<2g'^2$). 
%
%
%
%
We find numerically that $g'+{1\over 2} y g''(y_0)>0$ {\footnote{It is
not clear if we should ask for a minimum in the space $(u,v)$ as well
in which case the Hessian has to be considered.}} and hence we
conclude that the solution is a minimum.
In order to analyse the thermodynamic nature, ref. \cite{fgk} suggests
that one computes the Weinhold metric $W_{ab}=1/{(2\sqrt{S_{BH}})}
\partial_a \partial_b S_{BH}(u,v,y)$ with the $a,b$ belonging to the
space of scalar moduli. The positive definiteness of the Weinhold
metric is required for thermodynamic stability. If we restrict to the
space of scalar fields, our numerical results show that the metric is
positive definite. For the negative entropy solutions, the result is
interesting as well. For these solutions with
$(u_0,v_0,y_0)=(0.1472,0.6144,\pm 0.1213),$ and
$(u_0,v_0,y_0)=(0.0085,-0.0816,\pm 0.0923)$ we find that the entropy
is actually a maximum with respect to $y$. The implications of these
results are not entirely clear to us but probably imply that the first
solution is thermodynamically stable while the other two are
unstable. In passing we note that for Sen's heterotic fundamental
string with momentum and winding, the Weinhold metric in the space
$S,T$ is given by
$$
{\sqrt{8\pi}\over (nw)^{3/4}}\begin{pmatrix}{1\over 2} & 0 \\ 0 &
  w^2\end{pmatrix}\,, 
$$ 
and hence is also positive-definite. However we take note of the fact
that negative entropy solutions for higher derivative gravity have
been reported elsewhere \cite{ms} (see \cite{ross} also) as well
although a general consensus for the implications has not been
reached. At the same time we do not see how to fix the integration
constant which has been proposed to be the resolution to the negative
entropy ambiguity. The other possibility is that the negative entropy
solutions could be unphysical and cease to be solutions when all the
quantum corrections are included.\footnote{We thank Rob Myers for
suggesting this possibility.}

\section{Discussion}

In this paper, we have considered the potential near horizon
geometries of extremal small black-holes in type IIA made from a large
number of D0-branes in 10-dimensions. We demonstrated using Sen's
entropy function formalism and making use of the 11-dimensional $R^4$
corrections to supergravity that the horizon becomes finite in higher
derivative gravity theory. The corresponding Wald's entropy is
proportional to the square-root of the electric charge of the RR
1-form of type IIA. As is usually the case with small black holes
there is no guarantee that the solution found at a finite order in a
derivative expansion survives further corrections. However based on
the microstate counting in the introduction we suggest that it
should. We will now discuss some important consequences of our
analysis and outstanding issues.

\vskip .4cm
\noindent{\bf \underline{The sub-leading corrections:}} Let us
consider the $R^4$ terms at the tree level in type IIA. Since we have
\begin{equation}
\frac{\alpha'^3}{\kappa_{10}^2}
\rightarrow \frac{\alpha'^3}{R_{11} \, \kappa_{10}^2} =
\frac{1}{R_{11}^3},
\end{equation}
the tree-level $R^4$ term translates into ${\cal O}(\frac{1}{l_{11}^3}
(\frac{l_{11}}{R_{11}})^3)$. Using $(\frac{l_{11}}{R_{11}})^3 =
\frac{1}{s^3}$ the correction adds to the 11-dimensional $R^4$ term as
a sub-leading one. Putting things together the action would take the
general form:
\begin{equation}
f(u,v,e,s) = s \left[ g (u,v,y) + \frac{1}{s^3} \tilde g(u,v,y) \right]
\end{equation}
Now going through the exercise of section 3 we get 
\begin{eqnarray}
S_{BH} &=&  2\pi \, \left[ \alpha \sqrt{|q|} + \frac{\beta}{|q|} +
  \cdots 
\right]. 
\end{eqnarray}
So we see that the string tree-level $R^4$ terms in type IIA seen from
M-theory can only contribute to the sub-leading corrections to the
entropy (in the large-$q$ limit).

\vskip .4cm
\noindent{\bf \underline{Other M-theory higher derivative terms:}}
Let us analyse the effect of other higher derivative terms in
M-theory. First let us assume that the leading entropy result is
$\sqrt{q}$ and subsequent terms will be sub-leading. It is clear from
our analysis that the leading term in the entropy function should be
of the form $s g(u,v,e s)$ and any factor of $s$ with higher powers
will naively lead to higher powers of $q$. Suppose that in the IIA
action there are terms of the form $\sqrt{g} R^m e^{2 n\phi}$. This
will lead to $s^{5-m+3n}$ as the $s$ dependent term multiplying
$g(u,v,e s)$. This leads to the constraint
\be
n\leq [{m-4\over 3}]\,,
\ee
where $[x]$ denotes the integer part of $x$. This formula tells us
that the maximum string loop at $O(R^m)$ is $1+n$. Suppose $m=4$. Then
this immediately leads to the well-known result that the maximum
string loop correcting the $R^4$ term is 1 and all higher string loops
should vanish. At $R^7$ this leads to the prediction that only two
loops and below will contribute. However this result seems to be in
contradiction to what the authors of \cite{gv2005} have found which
suggests that there should also be a contribution at three loops. What
is the resolution to this apparent puzzle?

The $\sqrt{q}$ dependence from the counting is only the leading order
result in the expansion. The sub-leading terms take on the form
\be
{\rm constant}+ \log q+{1\over \sqrt{q}}+\cdots\,.
\ee
We have found that the tree level IIA $R^4$ term gives rise to $1/q$
correction to the entropy formula. We still have to account for the
other terms in the above equation. A simple power counting argument
tells us that if there was a term $\sqrt{g} R^5$ in the IIA effective
action this would lead to the constant term. $\sqrt{g} R^6$ would lead
to $1/\sqrt{q}$. What about the $\log q$ dependence?

We conjecture that all the terms that naively violate the $s g(u,v,e
s)$ scaling property by introducing a higher power of $s$ as the
prefactor will resum into a sub-leading $\log q$ dependence. Although
it is not clear how this is going to happen, it is clear that this is
going to be a highly restrictive constraint and will help in verifying
the structure of the M-theory effective action. {\footnote{In passing
we note, that if the entropy function had the form $s^n g(u,v, es)$,
then the entropy would behave like $q^{n\over n+1}$ and for
$n\rightarrow \infty$, the entropy would go like $q$. }}

\vskip .4cm
\noindent {\bf \underline{The issue of supersymmetry:}} A very
important question that remains unanswered is if our $AdS_2 \times
S^8$ solution is supersymmetric or not. This is a hard question to
answer since the relevant higher derivative modifications of the
supersymmetric transformation rules are unknown. However we briefly
comment on both the scenarios:

\begin{enumerate}

\item If the solution we have found here describes a
non-supersymmetric configuration, this would be an example that Sen's
entropy function formalism gives sensible results for non-
supersymmetric black holes as well. In such a case our result that the
Sen's entropy computation and the statistical entropy scale the same
way with the charge must be because of the generalised attractor
mechanism of Sen \cite{entropyfn} now applied to the dilaton as
well.\footnote{We thank Ashoke Sen for suggesting this possibility.}
This means that the near horizon geometry does not depend on the
asymptotic value of the dilaton which decided the string coupling.

\item On the other hand the near horizon geometry and the potential
interpolating D0-brane small black hole solution could be in fact
supersymmetric. From the type IIA point of view the microstates should
be various ground states of the quantum mechanics of $N$
D0-branes. But there is a well-known conjecture that there is a unique
normalisable bound state \cite{ss,mns} of this quantum mechanics and
several non-bound states. All of these preserve 16
supersymmetries. The non-bound states can be seen as distinct
classical solutions only when we higgs the D0-brane quantum mechanics
which corresponds to spatially separating the $N$ D0-branes into
different subgroups. Such higgsing generically breaks the $SO(9)$
invariance of the quantum mechanics. If we try to restore the $SO(9)$
symmetry all such classical configurations collapse to a single one!
Let us list a few examples where similar microstates do exist as
distinct BPS states. In $AdS_5 \times S^5$ background of type IIB for
fixed total U(1) R-charge $J$ there are as many BPS states as the
number of partitions of $J$. These have been considered in the context
of small black holes recently in \cite{nvs} (see also
\cite{bbjs}). Similarly for a given DLCQ momentum $p^+$ of the
maximally supersymmetric type IIB pp-wave \cite{bmn, mrv} it is easy
to see that there is a degeneracy of BPS states given by the number of
its partitions. Finally the matrix model description of the DLCQ
compactification of the M-theory pp-wave also admits as many
supersymmetric vacua as the number of partitions of the light-cone
momentum \cite{bmn}.

\end{enumerate}

\noindent{But we leave the question of the supersymmetric nature of
our solution for future work.}

\vskip .2cm
\noindent{\underline{\bf The black hole and microstate geometries:}}
It will be important to find the interpolating solution between the
$AdS_2 \times S^8$ solution found here and the 10-dimensional flat
space ${\mathbb R}^{9,1}$. Another interesting question is what are
the microstate geometries \cite{mathur} of the extremal D0-brane
charged small black hole we considered here. It is plausible that the
well known M-waves \cite{hull} compactified on a circle are related to
these.

\vskip .4cm
\noindent{\bf \underline{Dual QM description:}} We have exhibited the
existence of a new vacuum solution of M-theory to the $R^4$ order of
the type $M_3 \times S^8$ where $M_3$ is locally $AdS_3$ with a null
1-dimensional boundary. From the type IIA point of view this geometry
is $AdS_2 \times S^8$ with RR 1-form electric field and a constant
dilaton. It would be interesting to see if there is a holographically
dual conformal quantum mechanics on the boundary. If our solution is
supersymmetric then there must be a way to account for the $\sqrt{|q|}$
scaling of the entropy from counting the number of chiral primaries in
the dual SCQM along the lines of \cite{cdkktp, townsend, strom}.

Finally it will be important to understand how to reconcile our
results with the conventional wisdom \cite{mathur} that one should not
associate a charge dependent entropy to single-charge systems like
strings with just winding number or momentum charge alone. We hope to
return to some of these questions in the future.

\section*{Acknowledgements}
We thank Keshav Dasgupta, Jerome Gauntlett, Gary Gibbons, Michael
Green, Sean Hartnoll, Gautam Mandal, Julian Sonner, Paul Townsend and
especially Rob Myers and Ashoke Sen for helpful discussions. AS
acknowledges support from PPARC and Gonville and Caius college,
Cambridge.

\end{document}